\documentclass{sf2a-conf2018}
\usepackage{graphicx}
\usepackage{hyperref}
\usepackage[]{natbib}  
\usepackage{epstopdf}
\usepackage{amsmath,amsfonts,amssymb}
\usepackage{textcomp,gensymb}

\def\BibTeX{{\rm B\kern-.05em{\sc i\kern-.025em b}\kern-.08em
    T\kern-.1667em\lower.7ex\hbox{E}\kern-.125emX}}
\bibpunct{(}{)}{;}{a}{}{,}  

\begin{document}

\TitreGlobal{SF2A 2018}


\title{F.S.I.: Flyby Scene Investigation}

\runningtitle{F.S.I.: Flyby Scene Investigation}

\author{N. Cuello$^{1,}$}\address{Instituto de Astrof\'isica, Pontificia Universidad Cat\'olica de Chile, Santiago, Chile,}\address{N\'ucleo Milenio de Formaci\'on Planetaria (NPF), Chile,} 
\author{G. Dipierro}\address{Department of Physics and Astronomy, University of Leicester, Leicester, LE1 7RH, United Kingdom,}
\author{D. Mentiplay}\address{Monash Centre for Astrophysics (MoCA) and School of Physics and Astronomy, Monash University, Clayton, Vic 3800, Australia,}
\author{D.~J. Price$^{4}$}
\author{C. Pinte$^{4}$}
\author{J. Cuadra$^{1,2}$} 
\author{G. Laibe}\address{Univ Lyon, Univ Lyon1, Ens de Lyon, CNRS, Centre de Recherche Astrophysique de Lyon UMR5574, F-69230, Saint-Genis-Laval, France,}
\author{F.~M\'enard}\address{Univ. Grenoble Alpes, CNRS, IPAG, F-38000 Grenoble, France,}
\author{P.P. Poblete$^{1,2}$}
\author{M. Montesinos$^{1,2}$}

\setcounter{page}{237}


\maketitle


\begin{abstract}
We present a methodology to interpret observations of protoplanetary discs where a flyby, also called a tidal encounter, is suspected. In case of a flyby, protoplanetary discs can be significantly disturbed. The resulting dynamical and kinematical signatures can last for several thousands of years after the flyby and hence deeply affect the evolution of the disc. These effects are stronger for closer encounters and more massive perturbers. For the very same flyby parameters, varying the inclination of the perturber's orbit produces a broad range of disc structures: spirals, bridges, warps and cavities. We study this kind of features both in the gas and in the dust for grains ranging from $1\,\mu$m to 10 cm in size. Interestingly, the dust exhibits a different dynamical behaviour compared to the gas because of gas-drag effects. Finally, flybys can also trigger high accretion events in the disc-hosting star, readily similar to FU~Orionis-type outbursts. All this information can be used to infer the flyby parameters from an incomplete set of observations at different wavelengths. Therefore, the main scope of our \textit{flyby scene investigation} (FSI) methodology is to help to interpret recent ``puzzling'' disc observations.
\end{abstract}



\section{Introduction}
\label{sec:intro}

Stars are born in molecular clouds, and planets in turn form in protoplanetary discs (PPDs) around young stars \citep{Armitage2011}. Therefore, in regions of high stellar density, discs do not evolve in isolation \citep{Pfalzner2013,Winter+2018b}. The turbulent evolution of the molecular cloud has dramatic effects on their protoplanetary discs as shown by \cite{Bate2018}. Moreover, the most recent observations of discs with {\sc sphere} and {\sc alma} have revealed spectacular disc structures: shadows \citep{stolker16a}, gaps and rings \citep{ALMA+2015}, spirals \citep{benisty15a}, warps \citep{Langlois+2018} and clumps \citep{casassus18a}. These challenge our understanding of disc evolution and planet formation. As a consequence, there is currently an active search for physical mechanisms able to create such features.

The presence of unseen planetary or low-mass stellar companions are among the favoured scenarios \citep[for example]{dong15a}. These bodies can be categorized into three main categories: \textit{inner}, \textit{embedded} and \textit{outer} companions. Inner companions are often invoked to explain the large cavities in transition discs and the asymmetries observed at the disc inner edge \citep[for HD~142527]{Price+2018}. Alternatively, embedded bodies efficiently carve deep gaps in the dust continuum and trigger spiral-wakes in the gas disc \citep[for HL~Tau]{Dipierro+2015}. Finally, outer companions located beyond the outer edge of the disc trigger ($m=2$) spirals and truncate the disc. In some cases, the companion has been directly imaged, as in HD~100453 \citep{Wagner+2015}. However, in other cases, the presence of hypothetical companions at large stellocentric distance are assumed, as for MWC~758 \citep{Dong+2018}.

There are also several mechanisms that do not require the presence of massive companions. For spirals alone, this includes gravitational instabilities \citep{kratter16a}; planetary companions embedded in the disc (cf. references above); external massive perturbers or stellar flybys \citep{Pfalzner2003,quillen05a,Dai+2015}; accretion from an external envelope \citep{Harsono+2011,Lesur+2015,Hennebelle+2016}; and asymmetric stellar illumination patterns \citep{montesinos16a,montesinos18a}.

In this work, we focus on the \textit{external massive perturber} scenario. We consider protoplanetary discs being perturbed by parabolic stellar flybys as in \cite{Clarke&Pringle1993} and \cite{XG2016}. Among all the possible inclinations, we focus on the two most likely\footnote{assuming that the distribution of the orbital inclination is uniform, i.e. a sort of ``flyby'' isotropy.}: retrograde and prograde orbits, both inclined with respect to the disc. All the results presented here are based on the recent study on the dynamical signatures of flybys in the gas and in the dust by \citet{Cuello+2018a}. In Section~\ref{sec:sims}, we present the simulations for inclined parabolic orbits. Then, in Section~\ref{sec:fsi}, we propose the FSI methodology that can be followed to interpret recent observations. We conclude in Section~\ref{sec:conclusions}.

\section{Flyby simulations}
\label{sec:sims}

\subsection{Numerical setup}
\label{sec:setup}

Our dust-gas hydrodynamics simulations were performed with the Smoothed-Particle Hydrodynamics (SPH) code {\sc phantom} \citep{Price+2017}. We performed two-fluid simulations in order to trace the evolution of both the gaseous and the dusty components of the disc \citep{Laibe&Price2012}. We consider an isothermal disc extending from 10 to 150 au around a $1 \, M_{\odot}$-star. Additionally, we initially place a $1 \, M_{\odot}$-stellar perturber at $1\,500$ au from the central star. We set its velocity in order to obtain a parabolic ($e=1$) orbit with a pericentre distance equal to $200$ au. $\beta$ refers to the angle between the orbital inclination and the initial disc plane. For a disc rotating in the anticlockwise direction, $-90\degree<\beta<90\degree$ and $90\degree<\beta<270\degree$ correspond to prograde and retrograde orbits (respectively). Here, we only consider $\beta=45\degree$ (inclined prograde) and $\beta=135\degree$ (inclined retrograde) orbits. For further details on the disc model and the initial conditions, refer to \cite{Cuello+2018a}.

\begin{figure}[ht!]
 \centering
 \includegraphics[width=0.89\textwidth,clip]{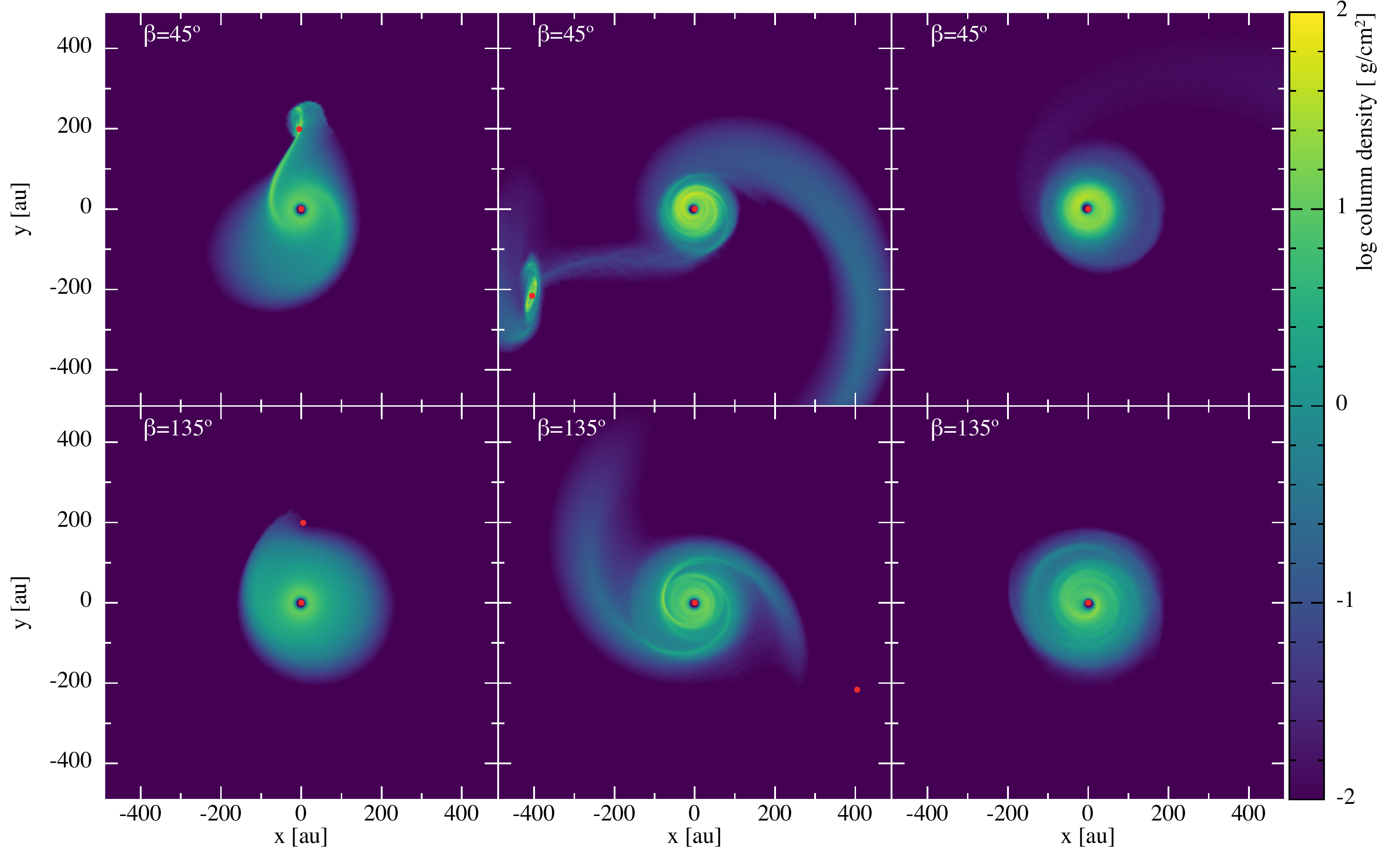}      
  \caption{Face-on view of the gas column density for $\beta45$ (top rows) and $\beta135$ (bottom rows). From left to right columns: $t=5\,400$ yr (pericentre), $t=5\,950$ yr, $t=8\,100$ yr. The disc rotation is anticlockwise. Sink particles (in red) are large for visualization purposes only. Spirals appear at (or shortly after) pericentre. The bridge between the two stars only appears for prograde configurations.}
  \label{Cuello1:fig1}
\end{figure}

\begin{figure}[ht!]
 \centering
 \includegraphics[width=0.89\textwidth,clip]{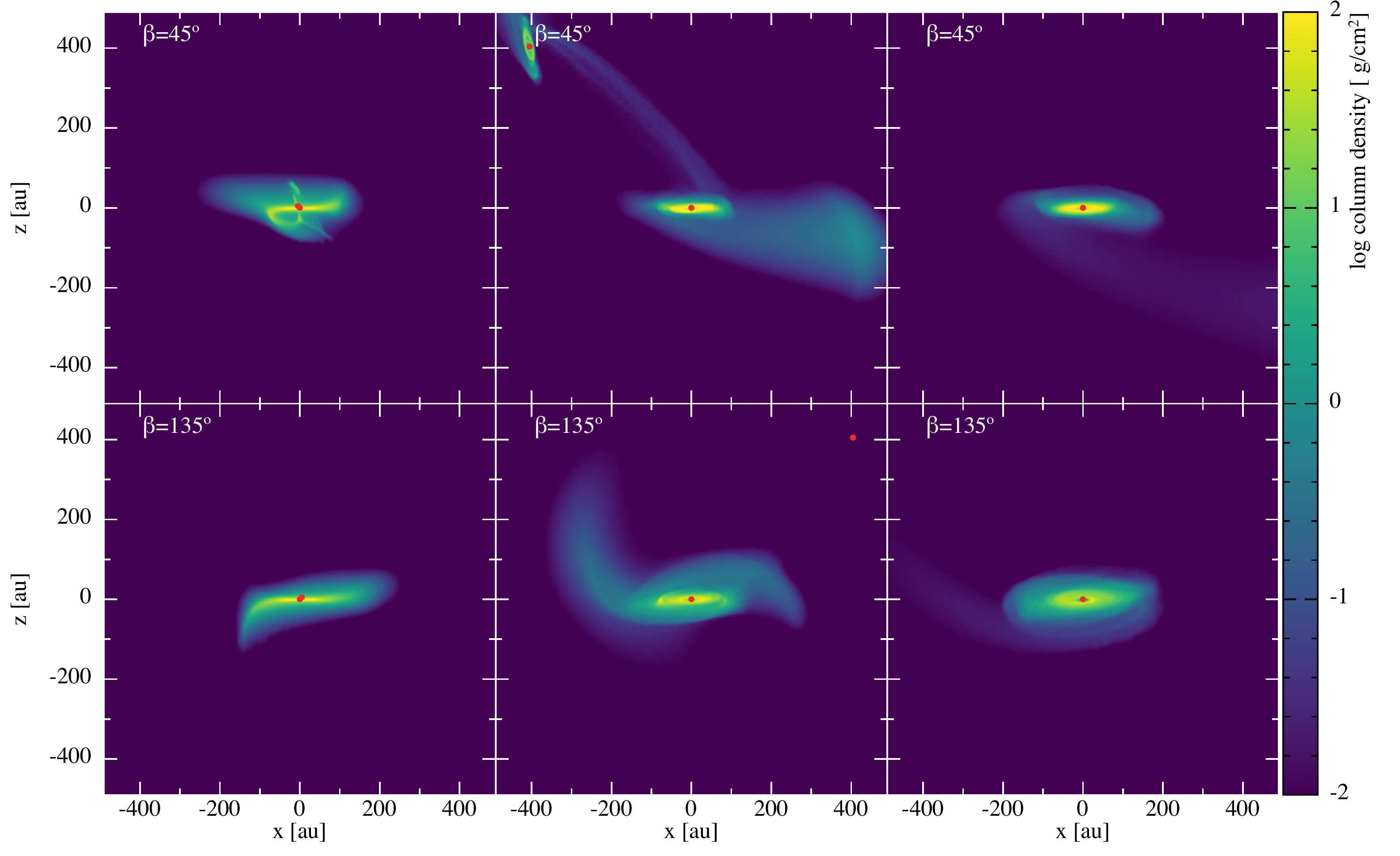}      
  \caption{Edge-on view of the gas column density for the same snapshots as Fig.~\ref{Cuello1:fig1}. Shortly after the passage at pericentre, the disc is warped (tilted and twisted).}
  \label{Cuello1:fig2}
\end{figure}

\begin{figure}[ht!]
 \centering
 \includegraphics[width=0.89\textwidth,clip]{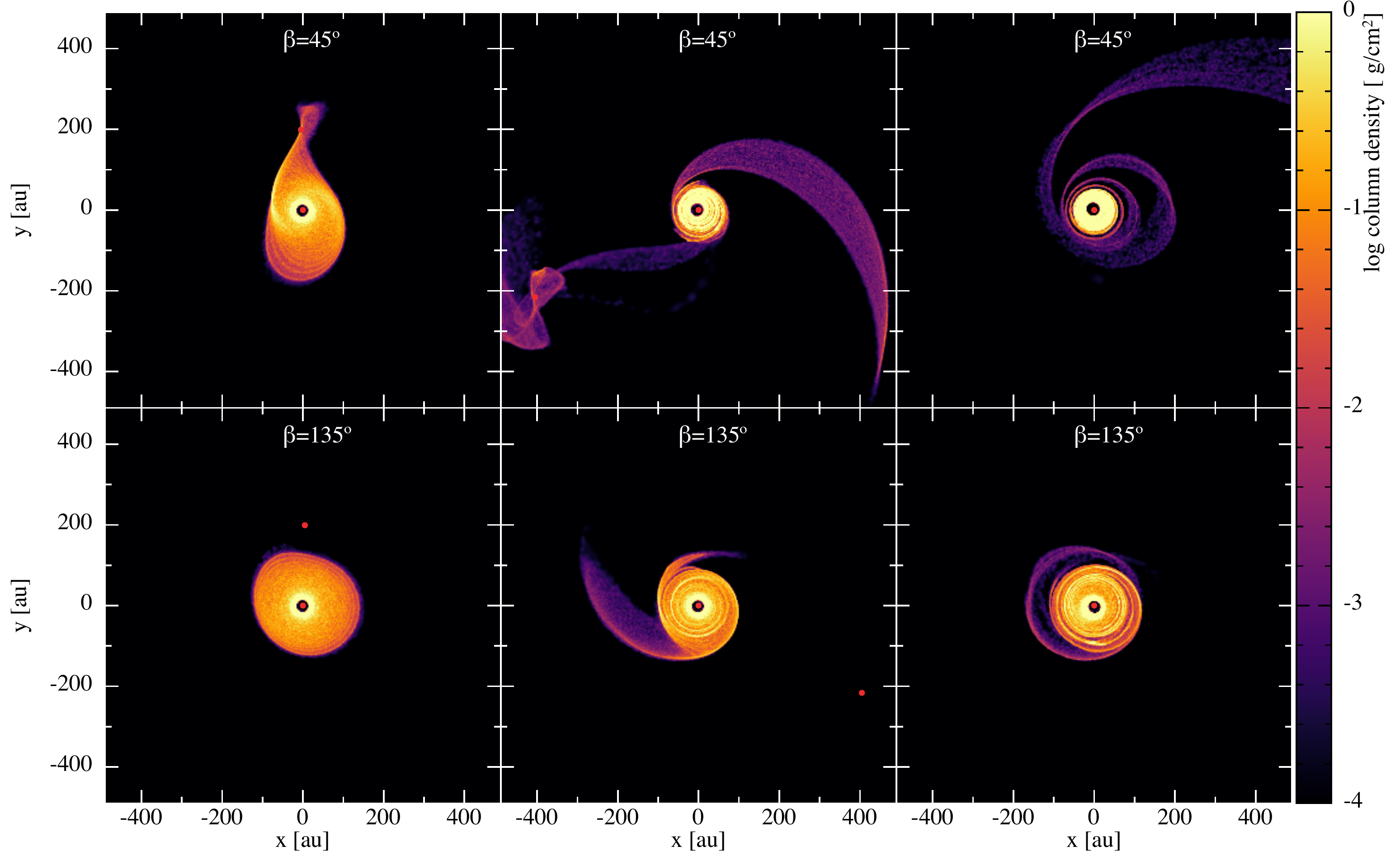}      
  \caption{Face-on view of the dust column density for$\beta45$ (top rows) and $\beta135$ (bottom rows). From left to right columns: $t=5\,400$ yr (pericentre), $t=5\,950$ yr, $t=8\,100$ yr. The disc rotation is anticlockwise. Sink particles (in red) are large for visualization purposes only. Spirals and bridges appear after the passage at pericentre as in Fig.~\ref{Cuello1:fig3}. The structures are sharper in the dust due to gas drag effects.}
  \label{Cuello1:fig3}
\end{figure}

\begin{figure}[ht!]
 \centering
 \includegraphics[width=0.89\textwidth,clip]{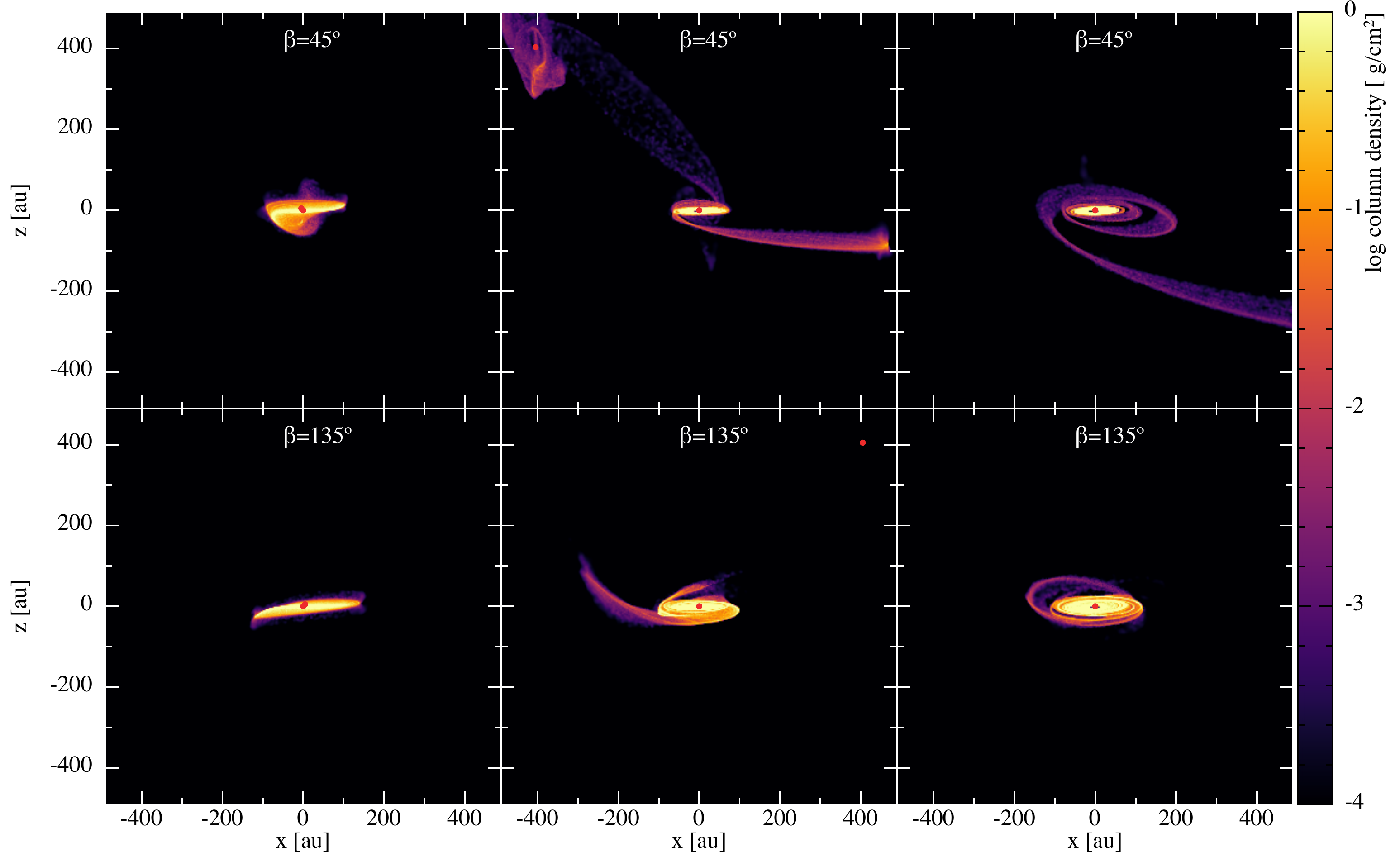}      
  \caption{Edge-on view of the dust column density for the same snapshots as Fig.~\ref{Cuello1:fig3}. The dusty discs are fairly thin in the vertical direction compared to the gas discs due to dust settling. Also, the discs are slightly tilted after the encounter.}
  \label{Cuello1:fig4}
\end{figure}

\subsection{Gas disc response to a flyby}
\label{eq:gas}

Figures~\ref{Cuello1:fig1} and \ref{Cuello1:fig2} show the column density of the gas for $\beta$45 (top rows) $\beta$135 (bottom rows) at different evolutionary stages during the flyby. In the $xy$-view (Figure~\ref{Cuello1:fig1}) we see that there are remarkable differences between the prograde and the retrograde configurations. In the former, shortly after the passage at pericentre, a ``bridge'' of material connects both stars. Moreover, disc material is captured around the perturbed (initially without a disc). Alternatively, no material is captured by the secondary for the retrograde orbit. Interestingly, in both cases, the arm on the disc side from which the perturber arrived is more extended. Finally, $2\,700$ yr after the passage at pericentre, the remaining disc is slightly eccentric. We also note that prograde perturbers truncate the disc more heavily compared to retrograde ones. 

In the $xz$-view (Figure~\ref{Cuello1:fig2}), we see that during and after the encounter disc material is lifted out of the $z=0$ plane. For example, the bridge connecting both stars for $\beta45$ is above the disc ($z>0$). Additionally, after several hundreds of years after the passage at pericentre, the tilt and the twist of the disc are modified compared to the initial values. For a more detailed study on this aspect, see \cite{Cuello+2018a}. Interestingly, all of these signatures are long-lived --- between several hundred years up to a few thousand years --- and can be used to infer the orbit of a suspected perturber, as shown in Section~\ref{sec:fsi}.

\subsection{Dust disc response to a flyby}
\label{eq:dust}

Figures~\ref{Cuello1:fig3} and \ref{Cuello1:fig4} show the column density of the mm-sized dust grains for $\beta$45 (top rows) and $\beta$135 (bottom rows) at different evolutionary stages during the flyby. We observe similar structure as in the gas disc. The main difference lies in the sharpness of the spirals and the radial extent of the disc. We also observe the ``bridge'' between both stars for $\beta$45. The dust being subject to gas drag experiences a rapid radial-drift, which explains the more radially-compact distributions of Figure~\ref{Cuello1:fig3}. The gas drag also leads to efficient dust settling, readily observed in the $xz$-views of Figure~\ref{Cuello1:fig4}. In \cite{Cuello+2018a} we show that gas drag plus the tidal effects are responsible for the dust trapping in the flyby-induced spirals. Again, prograde perturbers more severely affect the disc compared to retrograde ones.

The dust disc response depends strongly on the grain size, or equivalently to the local gas pressure in the disc. In \cite{Cuello+2018a} we show how grains with sizes ranging from $1\,\mu m$ up to $10$ cm exhibit different dynamical behaviours. Smaller grains are well-coupled to the gas and therefore mainly follow the gas distribution, while larger grains are decoupled from the gas and behave as test particles. In between --- mm- and cm-sized grains --- are marginally coupled to the gas and are efficiently trapped in the pressure maxima of the disc. Consequently, observations of discs being subject to a flyby are expected to vary with wavelength.

\section{Flyby Scene Investigation (FSI) methodology}
\label{sec:fsi}

\subsection{Flyby fingerprints}
\label{sec:fingerprints}

Based on the findings of Section~\ref{sec:sims}, we can establish a set of dynamical fingerprints that could indicate that a given disc is experiencing a flyby. Since this is based on the disc structure, the detection of the companion is irrelevant for this discussion. However, its detection can help to confirm the flyby hypothesis and narrow down the possible orbital configurations. Spirals, bridges, truncated discs, warps and non-coplanar material are indicators of an ongoing flyby. If a bridge of material --- plus perhaps some captured material --- is detected around the secondary, this allows one to distinguish between prograde and retrograde orbits. Additionally, in \cite{Cuello+2018a} we also show that the accretion rate onto the central star can significantly increase during prograde encounters. On the contrary, retrograde perturbers hardly modify the accretion rate. This statement is valid for non-penetrating encounters, i.e. when the pericentre is sufficiently large compared to the disc outer radius.

\begin{figure}[ht!]
 \centering
 \includegraphics[width=\textwidth,clip]{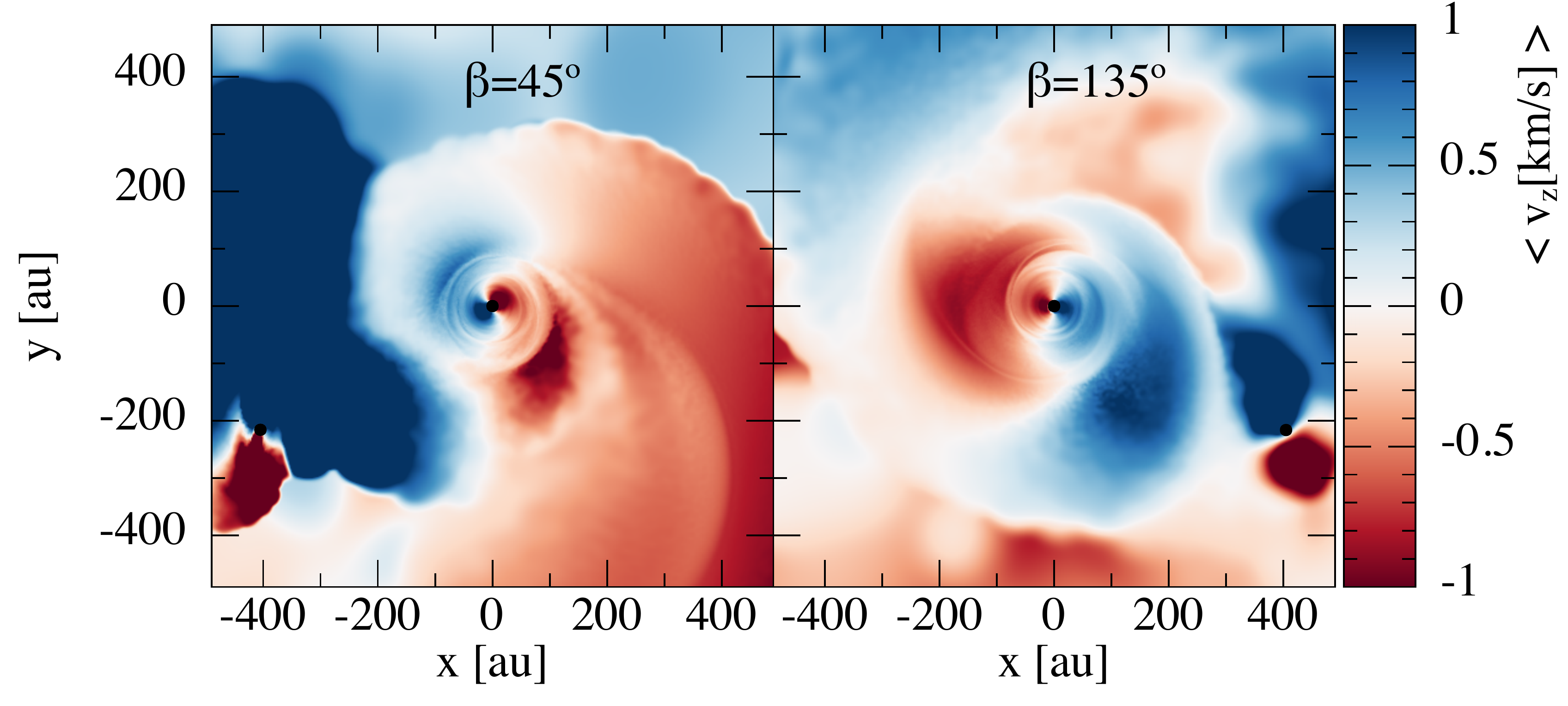}
  \caption{Mass weighted integral of $v_z$ along the line of sight, 450 yr after pericentre passage. $\beta=45\degree$ and $\beta=135\degree$ are shown on the left and right panels, respectively. The disc rotation is anticlockwise. The perturber is still in the ``field of observation''. The location of the red-shifted and blue-shifted regions of the disc provide information about the perturber’s orbit inclination. The stars are represented with black circles.}
  \label{Cuello1:fig5}
\end{figure}

If the flyby occurred long ago it is unlikely that the aforementioned disc features will be observed. However a warped (or even broken) disc can be the result of a flyby. Namely, massive perturbers and/or small pericentre distances lead to severe warping of the outer regions of the disc. Additionally, flyby-induced truncation may be responsible for unexpectedly compact discs. Lastly, if kinematic information of the system is available, it can be used to infer the orientation of the perturber's orbit. For example, Figure~\ref{Cuello1:fig5} shows the vertical velocity $\langle v_z \rangle$ integrated along the line of sight\footnote{defined as the mass-weighted integral: $\langle v_z \rangle \equiv \int \rho v_z {\rm d}z / \int \rho {\rm d} z$} 450 yr after the passage at pericentre. This quantity provides a dynamical signature comparable to the information obtained through CO lines observations. By comparing the kinematic fields of $\beta45$ to the ones in $\beta135$, we see that the sign of $\langle v_z \rangle$ is flipped. 

The ensemble of these fingerprints left in the disc by a single flyby, allow us to establish the ``culprit'''s orbital parameters. This can then be used by observers to interpret puzzling observations where a flyby is suspected. In the next subsection we present a practical case where we apply our proposed methodology.

\subsection{Practical case: retrograde or prograde?}
\label{sec:case}

For this exercise, let us assume that we observe a given disc --- located at 100 pc from the Earth --- with {\sc sphere} and {\sc alma} (continuum +  CO lines). This means that we have information at various wavelengths. We further assume that there is no companion detected in the $8 \times 8$ arcsec$^2$ field. In particular, scattered light traces the micron-sized grains, which are well-coupled to the gas; while the continuum emission (e.g. {\sc alma} band 7) corresponds to the thermal emission of grains of sizes ranging between 0.1 mm and 1 cm. Finally, the CO lines inform us about the kinematics of the disc. In other words, {\sc sphere} observations correspond to Figs.~\ref{Cuello1:fig1} and \ref{Cuello1:fig2}; while {\sc alma} observations are comparable to Figs.~\ref{Cuello1:fig3}, \ref{Cuello1:fig4} and \ref{Cuello1:fig5}. In the following, North is up and East to the left.

The careful analysis of the data reveals an almost face-on disc two prominent spirals arms in scattered light --- in the North and in the South --- with large pitch angles. Moreover, one of the Southern spiral arms is brighter compared to Northern one. On the contrary, the dust continuum exhibits a compact disc with only one faint and open spiral in the North. The different velocity maps --- obtained through CO lines --- show that the Eastern and Western sides are blue- and red-shifted, respectively. This suggests that the disc, despite being almost face-on, is warped. Given this set of observations, it is reasonable to suspect a flyby where the companion has already left the field.

Based on the flyby fingerprints presented in Section~\ref{sec:fingerprints}, we are prone to think that the hypothetical perturber was on a \textit{prograde inclined} orbit, i.e. a configuration comparable to $\beta45$. In this particular case, the faint open spiral in the continuum in the North is crucial to distinguish between prograde and retrograde configurations. In fact, if it were a retrograde encounter, then then prominent spiral would appear in the South. In case of strong H$_\alpha$ emission from the central star, this could indicate that there is an ongoing accretion event. In such case, the chances are high that the companion is on a prograde orbit \citep{Cuello+2018a}. Assuming that our conclusions are correct, then the ``culprit'' should be located to the East of the disc in the sky. Moreover, it is likely moving towards the observer --- in the reference frame of the system.

\section{Conclusions}
\label{sec:conclusions}

We have presented here the dynamical signatures left by inclined (prograde and retrograde) stellar flybys in the gas and in the dust of a protoplanetary disc. As shown in Section~\ref{sec:sims}, the tidal interaction creates spirals, bridges, warps and eccentric discs. Interestingly, some of these features are able to survive for several thousands of years. From the observational point of view, this means that the hypothetical perturber might have left the field. Therefore, stellar flybys seem a promising scenario to explain some of the recent observations of discs with prominent spirals where (despite an active search) no companions have been detected.

The FSI methodology detailed in Section~\ref{sec:fsi} allows to infer the inclination of the orbit based on a limited set of observations. In fact, retrograde and prograde orbits produce dramatically different structure in the disc (truncation, spirals, warps, accretion rates). This remarkable fact could help to hunt down ``stellar suspects'' (a.k.a. perturbers) at distances of several hundreds of au from the disc. If the culprit cannot be identified, this methodology at least provides useful information about the flyby that might have occurred in the past. The detailed observational diagnostics of flybys, i.e. the appearance of the disc at different wavelengths, will be the subject of a future investigation (Cuello et al., in prep.). Either way, stellar encounters might be more common than previously thought.

As Maggie Simpson once said: ``This is indeed a disturbing Universe''.

\begin{acknowledgements}
N.C. acknowledges financial support provided by FONDECYT grant 3170680. GD acknowledges financial support from the European Research Council (ERC) under the European Union's Horizon 2020 research and innovation programme (grant agreement No 681601). JC acknowledges financial support from the ICM (Iniciativa Cient\'ifica Milenio) via the N\'ucleo Milenio de Formaci\'on Planetaria grant, and from CONICYT project Basal AFB-170002.
The Geryon2 cluster housed at the Centro de Astro-Ingenier\'ia UC was used for the calculations performed in this paper. The BASAL PFB-06 CATA, Anillo ACT-86, FONDEQUIP AIC-57, and QUIMAL 130008 provided funding for several improvements to the Geryon/Geryon2 cluster. Part of this research used the ALICE High Performance Computing Facility at the University of Leicester. Some resources on ALICE form part of the DiRAC Facility jointly funded by STFC and the Large Facilities Capital Fund of BIS.
\end{acknowledgements}

\bibliographystyle{aa}
\bibliography{Cuello1}

\begin{thebibliography}{28}
\expandafter\ifx\csname natexlab\endcsname\relax\def\natexlab#1{#1}\fi

\bibitem[{{ALMA Partnership} {et~al.}(2015){ALMA Partnership}, {Brogan},
  {P{\'e}rez}, {Hunter}, {Dent}, {Hales}, {Hills}, {Corder}, {Fomalont},
  {Vlahakis}, {Asaki}, {Barkats}, {Hirota}, {Hodge}, {Impellizzeri}, {Kneissl},
  {Liuzzo}, {Lucas}, {Marcelino}, {Matsushita}, {Nakanishi}, {Phillips},
  {Richards}, {Toledo}, {Aladro}, {Broguiere}, {Cortes}, {Cortes}, {Espada},
  {Galarza}, {Garcia-Appadoo}, {Guzman-Ramirez}, {Humphreys}, {Jung}, {Kameno},
  {Laing}, {Leon}, {Marconi}, {Mignano}, {Nikolic}, {Nyman}, {Radiszcz},
  {Remijan}, {Rod{\'o}n}, {Sawada}, {Takahashi}, {Tilanus}, {Vila Vilaro},
  {Watson}, {Wiklind}, {Akiyama}, {Chapillon}, {de Gregorio-Monsalvo}, {Di
  Francesco}, {Gueth}, {Kawamura}, {Lee}, {Nguyen Luong}, {Mangum}, {Pietu},
  {Sanhueza}, {Saigo}, {Takakuwa}, {Ubach}, {van Kempen}, {Wootten},
  {Castro-Carrizo}, {Francke}, {Gallardo}, {Garcia}, {Gonzalez}, {Hill},
  {Kaminski}, {Kurono}, {Liu}, {Lopez}, {Morales}, {Plarre}, {Schieven},
  {Testi}, {Videla}, {Villard}, {Andreani}, {Hibbard}, \&
  {Tatematsu}}]{ALMA+2015}
{ALMA Partnership}, {Brogan}, C.~L., {P{\'e}rez}, L.~M., {et~al.} 2015, \apj,
  808, L3

\bibitem[{{Armitage}(2011)}]{Armitage2011}
{Armitage}, P.~J. 2011, \araa, 49, 195

\bibitem[{{Bate}(2018)}]{Bate2018}
{Bate}, M.~R. 2018, \mnras

\bibitem[{{Benisty} {et~al.}(2015){Benisty}, {Juhasz}, {Boccaletti},
  {Avenhaus}, {Milli}, {Thalmann}, {Dominik}, {Pinilla}, {Buenzli}, {Pohl},
  {Beuzit}, {Birnstiel}, {de Boer}, {Bonnefoy}, {Chauvin}, {Christiaens},
  {Garufi}, {Grady}, {Henning}, {Huelamo}, {Isella}, {Langlois}, {M{\'e}nard},
  {Mouillet}, {Olofsson}, {Pantin}, {Pinte}, \& {Pueyo}}]{benisty15a}
{Benisty}, M., {Juhasz}, A., {Boccaletti}, A., {et~al.} 2015, \aap, 578, L6

\bibitem[{{Casassus} {et~al.}(2018){Casassus}, {Avenhaus}, {P{\'e}rez},
  {Navarro}, {C{\'a}rcamo}, {Marino}, {Cieza}, {Quanz}, {Alarc{\'o}n}, {Zurlo},
  {Osses}, {Rannou}, {Rom{\'a}n}, \& {Barraza}}]{casassus18a}
{Casassus}, S., {Avenhaus}, H., {P{\'e}rez}, S., {et~al.} 2018, \mnras

\bibitem[{{Clarke} \& {Pringle}(1993)}]{Clarke&Pringle1993}
{Clarke}, C.~J. \& {Pringle}, J.~E. 1993, \mnras, 261, 190

\bibitem[{{Cuello} {et~al.}(2018){Cuello}, {Dipierro}, {Mentiplay}, {Price},
  {Pinte}, {Cuadra}, {Laibe}, F, P.P., \& M.}]{Cuello+2018a}
{Cuello}, N., {Dipierro}, G., {Mentiplay}, D., {et~al.} 2018, \mnras, (subm.)

\bibitem[{{Dai} {et~al.}(2015){Dai}, {Facchini}, {Clarke}, \&
  {Haworth}}]{Dai+2015}
{Dai}, F., {Facchini}, S., {Clarke}, C.~J., \& {Haworth}, T.~J. 2015, \mnras,
  449, 1996

\bibitem[{{Dipierro} {et~al.}(2015){Dipierro}, {Price}, {Laibe}, {Hirsh},
  {Cerioli}, \& {Lodato}}]{Dipierro+2015}
{Dipierro}, G., {Price}, D., {Laibe}, G., {et~al.} 2015, \mnras, 453, L73

\bibitem[{{Dong} {et~al.}(2018){Dong}, {Liu}, {Eisner}, {Andrews}, {Fung},
  {Zhu}, {Chiang}, {Hashimoto}, {Liu}, {Casassus}, {Esposito}, {Hasegawa},
  {Muto}, {Pavlyuchenkov}, {Wilner}, {Akiyama}, {Tamura}, \&
  {Wisniewski}}]{Dong+2018}
{Dong}, R., {Liu}, S.-y., {Eisner}, J., {et~al.} 2018, \apj, 860, 124

\bibitem[{{Dong} {et~al.}(2015){Dong}, {Zhu}, {Rafikov}, \& {Stone}}]{dong15a}
{Dong}, R., {Zhu}, Z., {Rafikov}, R.~R., \& {Stone}, J.~M. 2015, \apj, 809, L5

\bibitem[{{Harsono} {et~al.}(2011){Harsono}, {Alexander}, \&
  {Levin}}]{Harsono+2011}
{Harsono}, D., {Alexander}, R.~D., \& {Levin}, Y. 2011, \mnras, 413, 423

\bibitem[{{Hennebelle} {et~al.}(2016){Hennebelle}, {Lesur}, \&
  {Fromang}}]{Hennebelle+2016}
{Hennebelle}, P., {Lesur}, G., \& {Fromang}, S. 2016, \aap, 590, A22

\bibitem[{{Kratter} \& {Lodato}(2016)}]{kratter16a}
{Kratter}, K. \& {Lodato}, G. 2016, Annual Review of Astronomy and
  Astrophysics, 54, 271

\bibitem[{{Laibe} \& {Price}(2012)}]{Laibe&Price2012}
{Laibe}, G. \& {Price}, D.~J. 2012, \mnras, 420, 2345

\bibitem[{{Langlois} {et~al.}(2018){Langlois}, {Pohl}, {Lagrange}, {Maire},
  {Mesa}, {Boccaletti}, {Gratton}, {Denneulin}, {Klahr}, {Vigan}, {Benisty},
  {Dominik}, {Bonnefoy}, {Menard}, {Avenhaus}, {Cheetham}, {Van Boekel}, {de
  Boer}, {Chauvin}, {Desidera}, {Feldt}, {Galicher}, {Ginski}, {Girard},
  {Henning}, {Janson}, {Kopytova}, {Kral}, {Ligi}, {Messina}, {Peretti},
  {Pinte}, {Sissa}, {Stolker}, {Zurlo}, {Magnard}, {Blanchard}, {Buey},
  {Suarez}, {Cascone}, {Moller-Nilsson}, {Weber}, {Petit}, \&
  {Pragt}}]{Langlois+2018}
{Langlois}, M., {Pohl}, A., {Lagrange}, A.~M., {et~al.} 2018, \aap, 614, A88

\bibitem[{{Lesur} {et~al.}(2015){Lesur}, {Hennebelle}, \&
  {Fromang}}]{Lesur+2015}
{Lesur}, G., {Hennebelle}, P., \& {Fromang}, S. 2015, \aap, 582, L9

\bibitem[{{Montesinos} \& {Cuello}(2018)}]{montesinos18a}
{Montesinos}, M. \& {Cuello}, N. 2018, \mnras, 475, L35

\bibitem[{{Montesinos} {et~al.}(2016){Montesinos}, {Perez}, {Casassus},
  {Marino}, {Cuadra}, \& {Christiaens}}]{montesinos16a}
{Montesinos}, M., {Perez}, S., {Casassus}, S., {et~al.} 2016, \apj, 823, L8

\bibitem[{{Pfalzner}(2003)}]{Pfalzner2003}
{Pfalzner}, S. 2003, \apj, 592, 986

\bibitem[{{Pfalzner}(2013)}]{Pfalzner2013}
{Pfalzner}, S. 2013, \aap, 549, A82

\bibitem[{{Price} {et~al.}(2018{\natexlab{a}}){Price}, {Cuello}, {Pinte},
  {Mentiplay}, {Casassus}, {Christiaens}, {Kennedy}, {Cuadra}, {Perez},
  {Marino}, {Armitage}, {Zurlo}, {Juhasz}, {Ragusa}, {Laibe}, \&
  {Lodato}}]{Price+2018}
{Price}, D.~J., {Cuello}, N., {Pinte}, C., {et~al.} 2018{\natexlab{a}}, \mnras,
  624

\bibitem[{{Price} {et~al.}(2018{\natexlab{b}}){Price}, {Wurster}, {Tricco},
  {Nixon}, {Toupin}, {Pettitt}, {Chan}, {Mentiplay}, {Laibe}, {Glover},
  {Dobbs}, {Nealon}, {Liptai}, {Worpel}, {Bonnerot}, {Dipierro}, {Ballabio},
  {Ragusa}, {Federrath}, {Iaconi}, {Reichardt}, {Forgan}, {Hutchison},
  {Constantino}, {Ayliffe}, {Hirsh}, \& {Lodato}}]{Price+2017}
{Price}, D.~J., {Wurster}, J., {Tricco}, T.~S., {et~al.} 2018{\natexlab{b}},
  \pasa, 35, e031

\bibitem[{{Quillen} {et~al.}(2005){Quillen}, {Varni{\`e}re}, {Minchev}, \&
  {Frank}}]{quillen05a}
{Quillen}, A.~C., {Varni{\`e}re}, P., {Minchev}, I., \& {Frank}, A. 2005, \aj,
  129, 2481

\bibitem[{{Stolker} {et~al.}(2016){Stolker}, {Dominik}, {Avenhaus}, {Min}, {de
  Boer}, {Ginski}, {Schmid}, {Juhasz}, {Bazzon}, {Waters}, {Garufi},
  {Augereau}, {Benisty}, {Boccaletti}, {Henning}, {Langlois}, {Maire},
  {M{\'e}nard}, {Meyer}, {Pinte}, {Quanz}, {Thalmann}, {Beuzit}, {Carbillet},
  {Costille}, {Dohlen}, {Feldt}, {Gisler}, {Mouillet}, {Pavlov}, {Perret},
  {Petit}, {Pragt}, {Rochat}, {Roelfsema}, {Salasnich}, {Soenke}, \&
  {Wildi}}]{stolker16a}
{Stolker}, T., {Dominik}, C., {Avenhaus}, H., {et~al.} 2016, \aap, 595, A113

\bibitem[{{Wagner} {et~al.}(2015){Wagner}, {Apai}, {Kasper}, \&
  {Robberto}}]{Wagner+2015}
{Wagner}, K., {Apai}, D., {Kasper}, M., \& {Robberto}, M. 2015, \apj, 813, L2

\bibitem[{{Winter} {et~al.}(2018){Winter}, {Clarke}, {Rosotti}, {Ih},
  {Facchini}, \& {Haworth}}]{Winter+2018b}
{Winter}, A.~J., {Clarke}, C.~J., {Rosotti}, G., {et~al.} 2018, \mnras, 478,
  2700

\bibitem[{{Xiang-Gruess}(2016)}]{XG2016}
{Xiang-Gruess}, M. 2016, \mnras, 455, 3086

\end{thebibliography}

\end{document}